\documentclass[manuscript]{copernicus}

\usepackage{url} 
\usepackage{hyperref}
\usepackage{amsmath}

\usepackage{soul}
\newcommand{\ulmo}{{\sc Ulmo}}

\begin{document}

%
%

\title{AI based Out-Of-Distribution Analysis of Sea Surface Height Data}

%
%




\Author[1, $\star$]{Benjamin}{Pritikin}
\Author[2,3,4,5, $\star$]{J. Xavier}{Prochaska}

\affil[1]{Earth and Planetary Sciences Department,
University of California, Santa Cruz, CA, 95064, USA}
\affil[2]{Affiliate of the Department of Ocean Sciences, University of California, Santa Cruz, CA, 95064, USA}
\affil[3]{Department of Astronomy and Astrophysics, University of California, Santa Cruz, CA, 95064, USA}
\affil[4]{Kavli Institute for the Physics and Mathematics of the Universe (Kavli IPMU), 5-1-5 Kashiwanoha, Kashiwa, 277-8583, Japan}
\affil[5]{Simons Pivot Fellow}


\runningtitle{TEXT}

\runningauthor{TEXT}

\firstpage{1}

\maketitle

\nolinenumbers

\begin{abstract}

We performed Out-Of-Distribution (OOD) analysis of 7.8 million Sea Surface Topography Merged Altimeter L4 cdr grid cutouts in an effort to identify rare 
(possibly unknown) physical phenomenon sea surface height (SSH) data. 
The algorithm used for the project is \ulmo\ which is a probabilistic autoencoder (PAE),
originally developed for sea surface temperature data. 
A PAE is made of an autoencoder for taking the extracted images and encoding them into a latent representation of the data, and a normalizing flow which takes the encoding and maps it to a normal distribution for probabilistic interpretation. 
A Log-Likelihood (LL) value for each cutout was calculated 
from this normal distribution and
we defined the images with the lowest 0.1 percentile of LL values
as anomalies. 
\ulmo\ successfully identifies outliers and distinguishes the ocean's most dynamic regions being Western boundary currents. 

\end{abstract}

\section*{Plain Language Summary}
Artificial Intelligence (AI) was used to analyze sea surface height data to find outliers in the dataset. The AI, \ulmo, consistently found the most extreme regions of the ocean independently. \ulmo\ found that the most extreme regions of the ocean are the warm fast currents on the western edge of the ocean basins known as Western boundary currents.

%
%

%


%
%
%
%

\section{Introduction}
This project is a continuation of our team's efforts to apply
machine learning (ML) algorithms to large remote sensing
datasets.  
In \cite{prochaska2021deep}, we
analyzed sea surface temperature (SST) to perform
an out-of-distribution (OOD) analysis. 
For this senior thesis, sea surface height observations (SSH) are the focus.

\begin{figure}[ht]
\noindent\includegraphics[width=\textwidth]{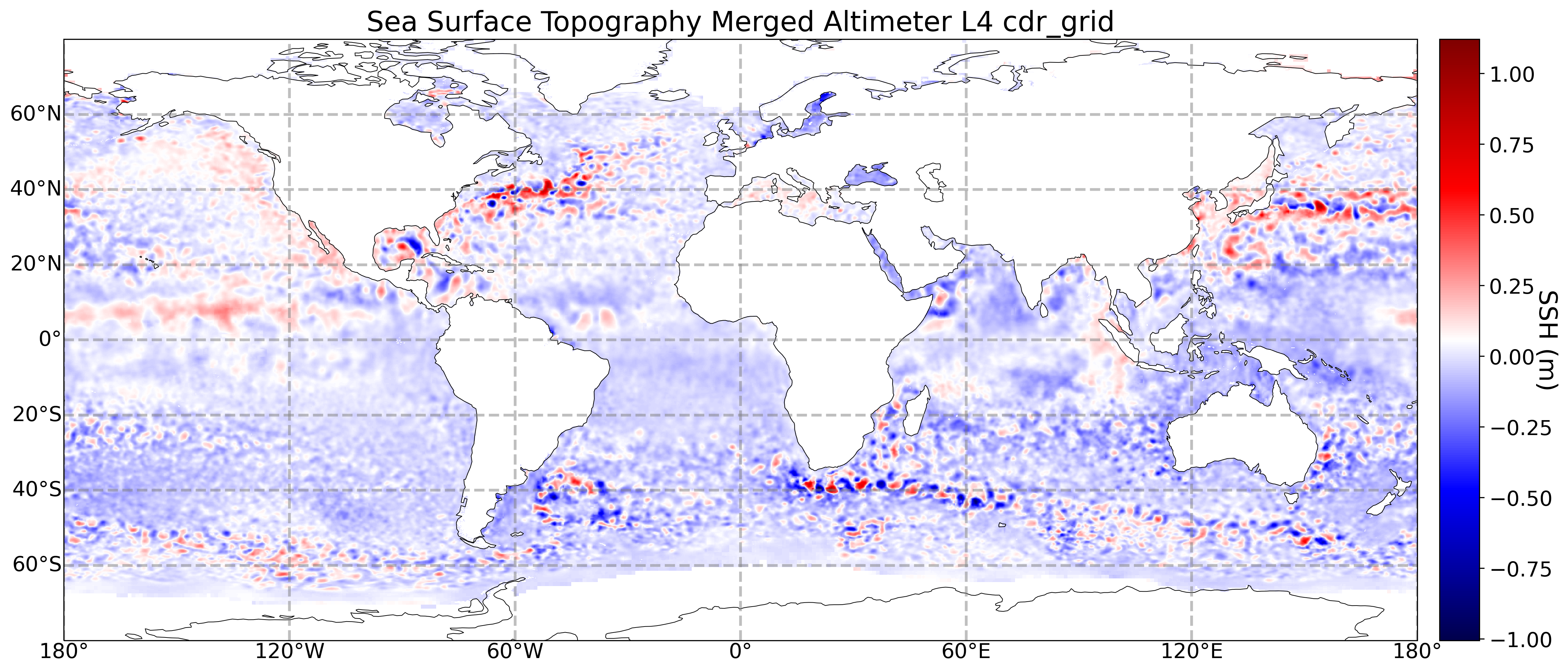}
\caption{Depicted is the first of 1922 files within the Sea Surface Topography Merged Altimeter L4 (Level four) cdr (Climate Data Record) grid courtesy of NASA JPL (Jet Propulsion Laboratory). This time slice is October 2, 1992. The distributed data ranges between about $80^\circ$S and $80^\circ$N. The dataset is made of altimetry data from TOPEX/Poseidon, Jason-1, Jason-2, Jason-3, and depending on the date ERS-1, ERS-2, Envisat, SARAL-AltiKa, and CRyosat-2 satellites.} 
\label{fig:1}
\end{figure}

SSH data is also known as surface ocean topography, and not to be confused with bathymetry which is the topography of the sea floor. 
A single day of the Level~4 product from  the Sea Surface Topography Merged 
Altimeter is
plotted in Figure~\ref{fig:1}. 
The SSH values range from $\approx  -1.5$ to 1.5~meters above or below the 
geoid. The geoid is where the sea-level would be in the absence of all variables except for gravity.  This includes effects related to bathymetry. 
SSH is affected by a variety of factors like tides, wind, temperature, pressure etc. SSH data is important because with it we can see how pressure gradients are manifested in the surface ocean \citep{zlotnicki1983oceanographic}. 
From these large-scale pressure gradients we may explain basin-scale gyres. 
We can also see from SSH data where up-welling or down-welling is occurring which is important for biogeochemical processes. We also observe mesoscale eddies which as distinct
features in SSH. 
Mesoscale processes are processes that have a spatial range of 50-500 km; however mesoscale eddies range in scale of tens to hundreds of kilometers \citep{fu2010eddy}.

\subsection{Motivations}
\subsubsection{Machine Learning}

The motivation for this project are two-fold.  The first is to 
explore the value of ML methods, which are powerful yet underutilized in the 
Earth and Marine sciences. ML has the potential to make significant breakthroughs in the Ocean sciences, but its potential has not yet been fully realized \citep{thessen2016adoption}.

In addition, ML methods thrive on large datasets of complex and multi-dimensional data; 
this gives ML techniques an advantage over standard statistical evaluation. This makes Earth and Marine science datasets perfect for ML applications. Many of these have continuously 
collected data for decades with high resolution and are open-source.

\subsubsection{Unknown Oceanography}

The second motivation for this project is the potential discovery of unknown physical phenomena. The type of ML strategy used for this project is unsupervised learning (USL). USL entitles training the machine on data that is neither labeled nor classified. It is up to the machine to find similarities, differences, and patterns that can be applied to the unsorted data without guidance \citep{ahmad2019machine}. A key advantage of this ML approach is that the machine may find phenomena that humans might
otherwise ignore because it has no predisposition or 
because of the overall size of the dataset.

USL is commonly used for tasks including clustering, dimensionality reduction, and anomaly detection \citep{ahmad2019machine}. For this project we are trying to identify extremes in SSH,
such that  anomaly detection is the task at hand. For the rest of this paper, anomaly detection will be referred to as out-of-distribution (OOD) detection. The distribution in OOD refers 
to a Gaussian (normal) distribution so if it is OOD than it is an extreme or an anomaly \citep{devries2018learning}.



\section{Data and Methods}

\subsection{Data}


 
 SSH today is measured using satellite altimetry. Altimeters are a radar based instrument for measuring the distance between a surface like the ocean and the satellite. 
 Satellite altimeters work by measuring the time it takes a radar pulse from the satellite to move between the satellite to the ocean and back to the satellite. The longer it takes the radar pulse to make the round trip, the farther away the surface 
 \citep{calman1987introduction}. 
 

\begin{figure}[h!]
\noindent\includegraphics[width=\textwidth]{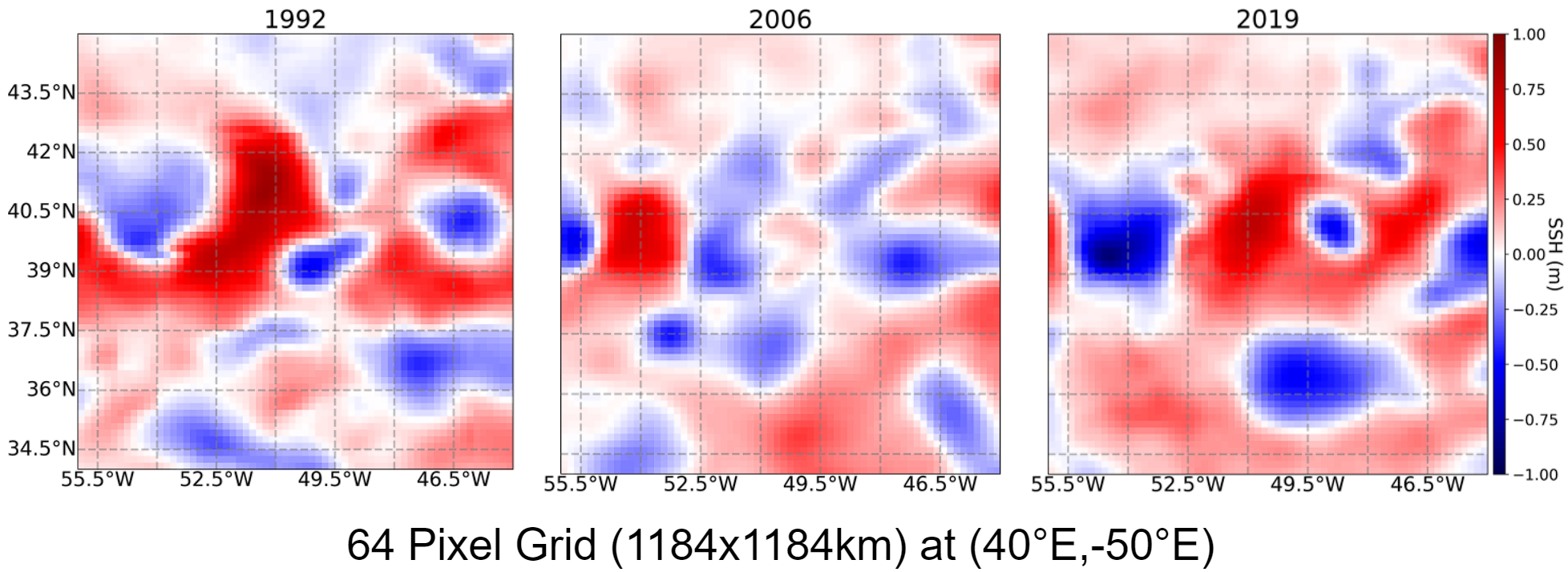}
\caption{Examples of 64$^2$ pixel grids at three distinct times.
This shows that despite not every satellite in the conglomerate being up for the entire duration of the dataset and not all satellites having equal resolution, 
the data does not show gross changes 
in resolution at any point in the dataset.} 
\label{fig:2}
\end{figure}

 There are several SSH datasets readily available and open source. These are made available by agencies like NOAA (National Ocean and Atmospheric Administration), NASA (National Aeronautics and Space Administration), and NCAR (National Center for Atmospheric Research). Most of these datasets are derived from one or several satellite altimeters. This means that a lot of these datasets are limited in temporal range or, if they have a unique orbit, spatially. The dataset chosen for this project is a conglomerate of at most eight satellite altimeters.
 
 
 The data downloaded for this project is the Sea Surface Topography Merged Altimeter L4 (Level four) cdr (Climate Data Record) grid courtesy of NASA JPL (Jet Propulsion Laboratory). 
 Data was downloaded using OPeNDAP URLs: \url{https://opendap.jpl.nasa.gov/opendap/SeaSurfaceTopography/merged_alt/L4/cdr_grid/contents.html}. 
The data paths all consist of a base path, the date of the data collection, and the file extension. To download the data, a date generator was made to create a list of dates to be added to the base URL and file extension to create the data path used to download with the unix executable {\tt wget} in a source file. 

For this project, a level four (L4) data product was used. Satellite data can undergo several levels of processing. An L4 data product is either model output or results from analysis of lower-level data; in this case it is the latter. The data used is a collection of level two (L2) data swaths from several satellite altimeters including TOPEX/Poseidon, Jason-1, Jason-2, Jason-3, and depending on the date ERS-1, ERS-2, Envisat, SARAL-AltiKa, and CRyosat-2. 
L2 data products are derived geophysical variables created from L1 data at the same resolution as the L1 data. L1a data products are telemetry data that has been extracted but not decommutated and formatted into time-sequenced datasets for further processing.

Datasets like this are only possible because of satellite oceanography and remote sensing techniques. Before the rise of satellites, obtaining SSH data required 
a buoy or  some other in-situ method would be necessary which limits spatial and temporal resolution. Remote sensing is the gathering of information without physically being there to measure it which makes large, open-ocean datasets the perfect target. With satellites we 
view the whole ocean all the time. 

The data has a resolution of $1/6^\circ$ which is approximately 18.5 km. The date given in the data is the center of the 5 day window beginning in October 2, 1992 and ending in January 24, 2019 at 12 pm. The first file in the dataset can be seen in figure~\ref{fig:1}. The distributed data ranges between approximately
$80^\circ$S and $80^\circ$N. There are 1,922 netCDF4 files each about 16.6 MB. All files together are about 30GB pre-extraction. The data ranges over time and not every satellite in the conglomerate was up for the entire period or at the same resolution as other satellites. D
espite this the data does exhibit gross changes 
in resolution in the period, as shown in figure~\ref{fig:2}.
 

\begin{figure}[h!]
\noindent\includegraphics[width=\textwidth]{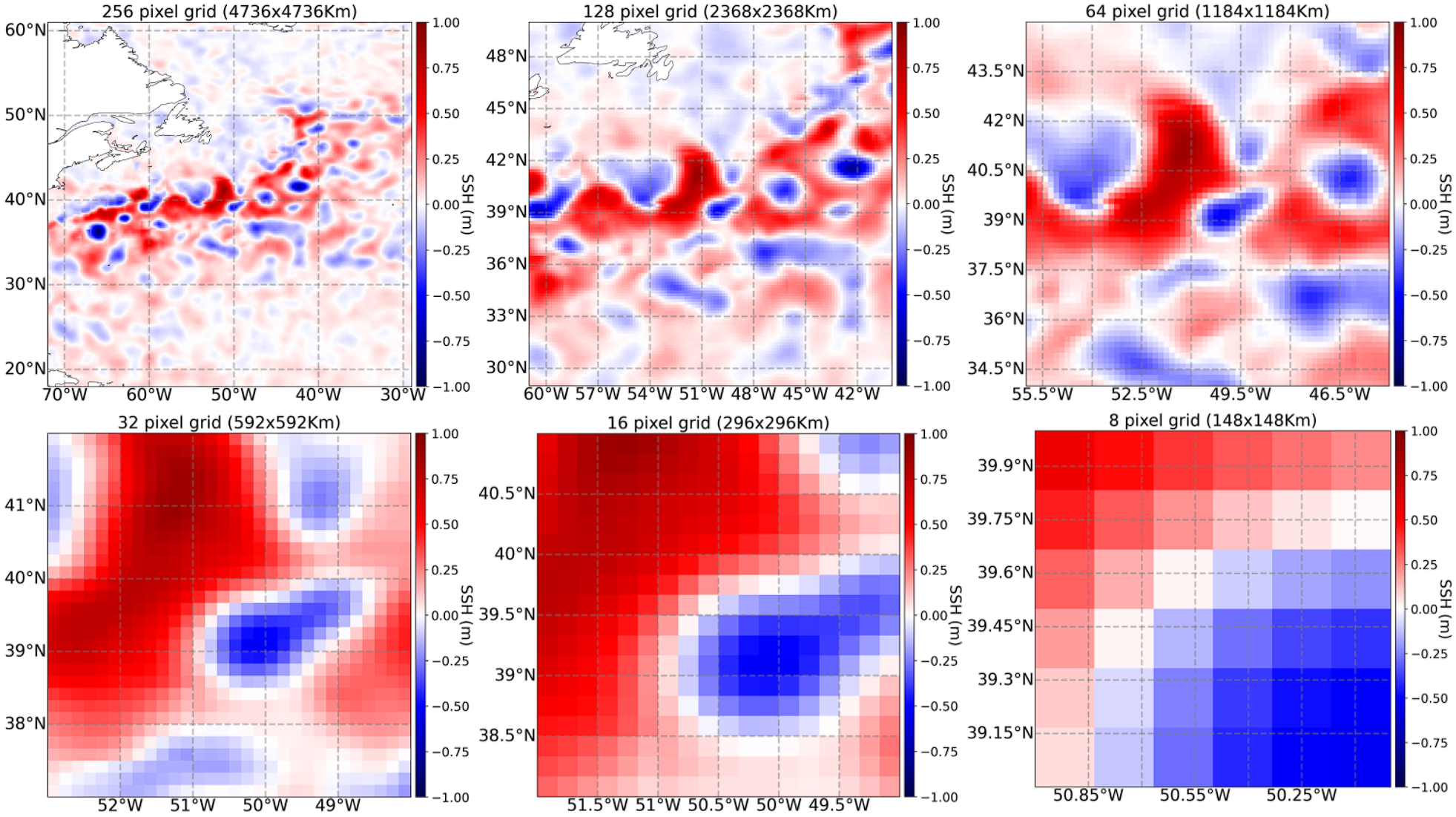}
\caption{Examples of different sized pixel grid cut-outs in decreasing scale. All images are centered on 40$^\circ$N,-50$^\circ$E in the year 2006. This point was chosen because it is far enough from land but is still within the Gulf Stream so it is rich in SSH features. 
The $8\times 8$~pixel cutout clearly has too few pixels to examine
meaningful physical features.
The $256^2$ pixel cutout, meanwhile, may key on basin-wide features.
By inspection, we down-selected to 
the 32$^2$ or 64$^2$ pixel cutouts.} 
\label{fig:3}
\end{figure}

\subsection{Extraction and Pre-Processing}

One of the first decisions was the scale on which to search for anamolies.
Consider the trade-offs between
32$\times$32 (592$\times$592 km) or 64$\times$64 (1184$\times$1184 km) pixel grids. 
Examples of varying pixel grid size can be seen in figure ~\ref{fig:3}. 
In the 32$\times$32 pixel grid scenario there would be plenty of data to train and validate the model on but each 
cutout would be so small that there may not be enough 
field-of-view to detect features like mesoscale eddies which can range from 10s to 100s of kilometers. 
In the 64$\times$64 pixel grid scenario, there would be sufficient
field-of-view for features to be visible, 
but there would be less data to evaluate. 
For the $32\times 32$ pixel grid there are 3462 cutouts per time slice
where we allow for 25\%\ overlap.
For the $64 \times 64$ pixel grid there are 617 cutouts per time slice. That is 6,653,964 and 320,974 total cutouts for 1922 time slices for $32^2$ and $64^2$ pixel grids respectively.

In the end, the $32^2$ pixel grid was chosen because it has an appropriate balance between having enough features per extraction, and because 
smaller cutouts would result in a larger dataset for training and validation. For the current project, there was not enough time to do both $32^2$ and $64^2$ pixel grids due to time constraints; however, this may be continued in a future project.

\begin{figure}[h!]
\noindent\includegraphics[width=\textwidth]{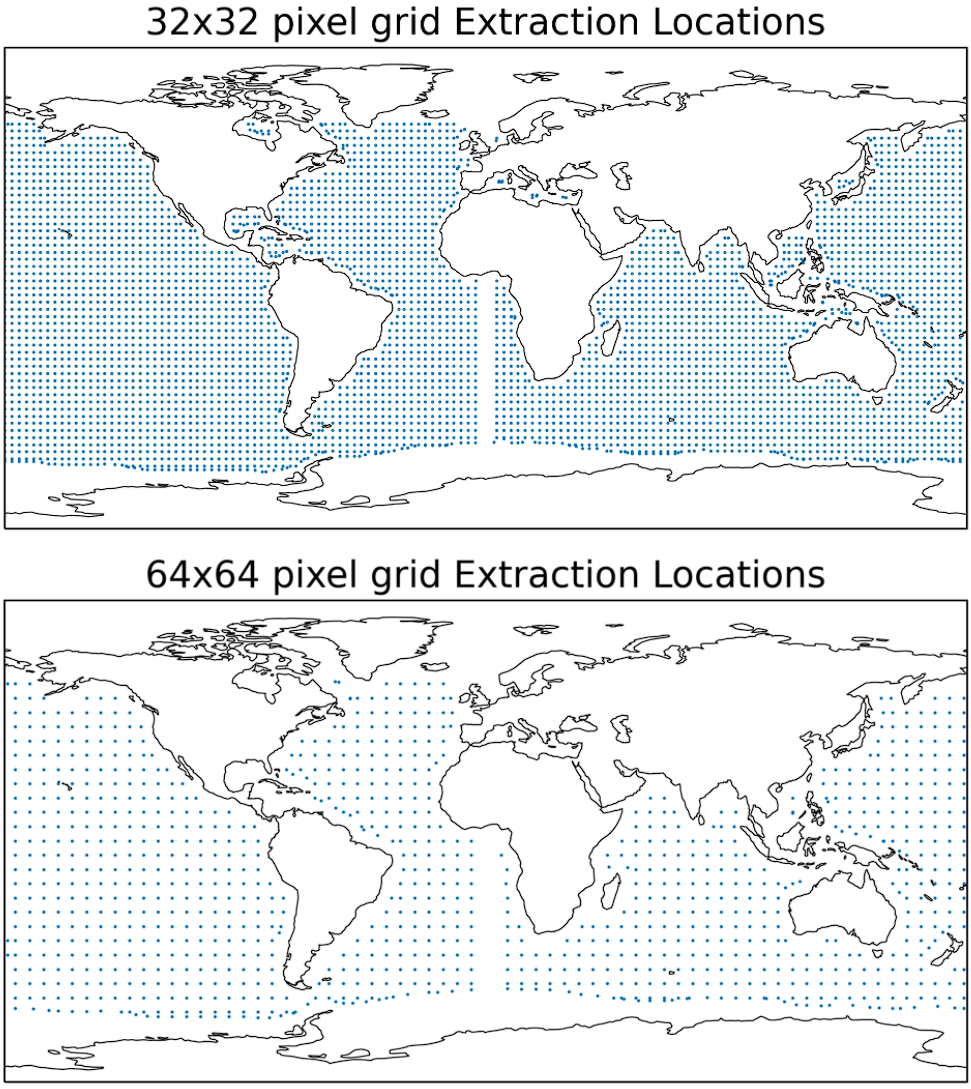}
\caption{The top plot is the location of every cutout at the $32^2$ pixel grid scale. The top plot is the location of every cutout at the $64^2$ pixel grid scale. There is a gap at the Prime Meridian because that is technically the spatial edge of the dataset. The extraction method used avoids having cutouts near islands and other borders.} 
\label{fig:4}
\end{figure}

Data was extracted with 25$\%$ overlap; this means that for every cutout there are four more cutouts that overlap it centered on the corners of the original.
The points of each cutout can be seen in figure~\ref{fig:4}.

\begin{figure}[h!]
\noindent\includegraphics[width=\textwidth]{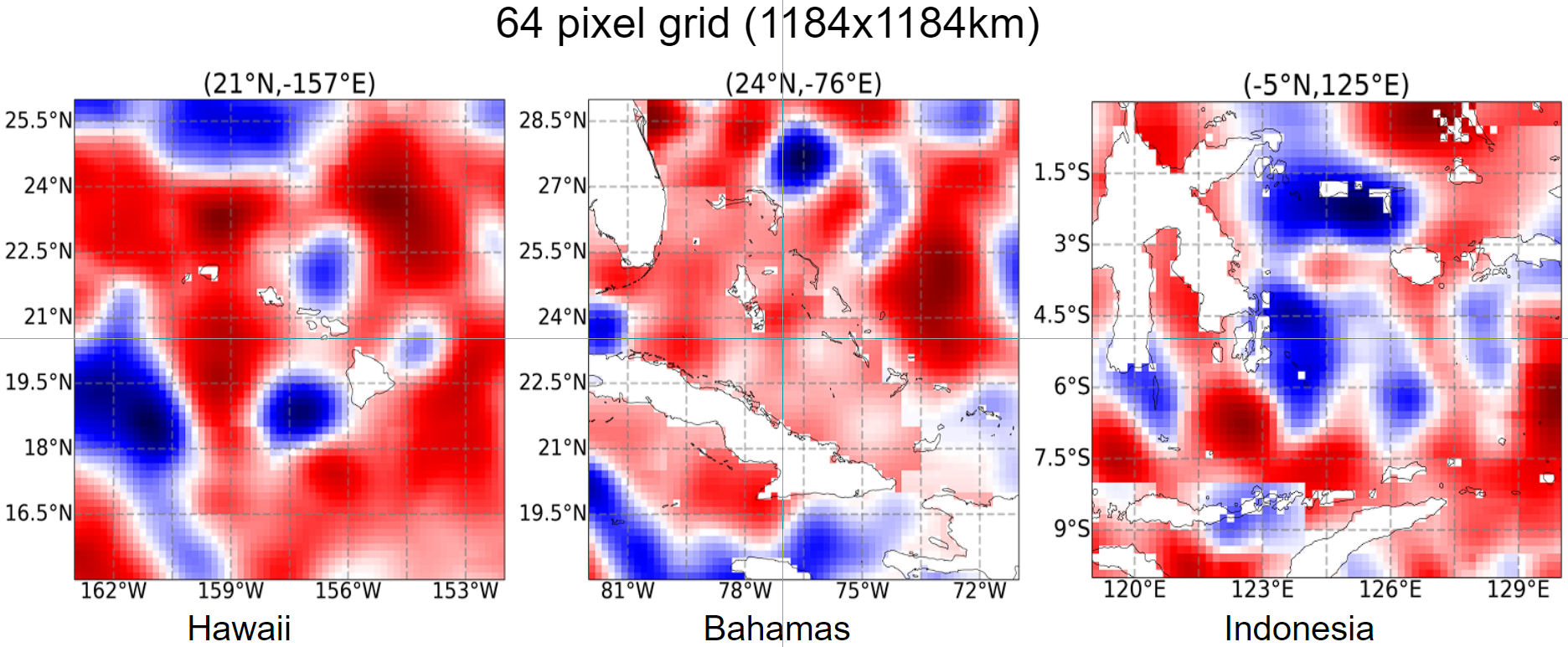}
\caption{Examples of regions with small islands to demonstrate what has been in-painted. These plots are from the first October 2, 1992 time slice. The plots are all 64$\times$64 (1184$\times$1184 km) pixel grids. All islands and land masses smaller than a pixel, 18.5 km, are already painted over. Note the stepping in the Bahamas and in Indonesia; this happens where the sea is shallow.} 
\label{fig:land}
\end{figure}

While the SSH observations are not affected by clouds,
the dataset does ignore land.
We considered
in-painting small islands to remove gaps in the data. In-painting is filling in missing the data by using an algorithm to replace the missing data with estimated values from nearby points. As seen in Figure~\ref{fig:land}, any land masses smaller than one pixel have already been painted over; this means that any land smaller than a 18.5km$^2$ are not in the dataset and have been replaced with estimated SSH values.  One also notices significant stepping in regions where the sea is shallow like in the Bahamas by 24.12$^\circ$N, -75.00$^\circ$E. 

\subsection{The PAE}


\ulmo\ is a machine learning algorithm developed by Prochaska et al. \ulmo\ is a probabilistic autoencoder (PAE) which is a likelihood-based generative model that combines two deep learning modules: an autoencoder and a normalizing flow \citep{prochaska2021deep,bohm2020probabilistic}. 

\subsubsection{Autoencoder}

One half of the PAE is an autoencoder. An autoencoder is a type of unsupervised artificial neural network (ANN) that takes unlabeled extracted data and encodes it into a latent representation of the original data. The autoencoder does two things: 
 (1) it makes an encoding and assigns each data a latent vector and 
 (2) the decoder converts the latent vector back into an approximation of the original image. 
The autoencoder takes in all the data and from that learns a mathematical function that can be used to describe all the images in the dataset, this function is the encoder. 
A latent vector is a compressed 128-dimensional representation of the original data. 
The latent vectors are what is put into the decoder to remake the original image. 
Latent vectors are reduced representations of the data,
effectively coordinates in a complex latent 
space\citep{sugiyama1965distribution}. 
The autoencoder
then refines and validates the data by remaking the input from the encoding 
In principle, one may use the decoder to generate random,
realistic data.

\begin{figure}[h!]
\noindent\includegraphics[width=\textwidth]{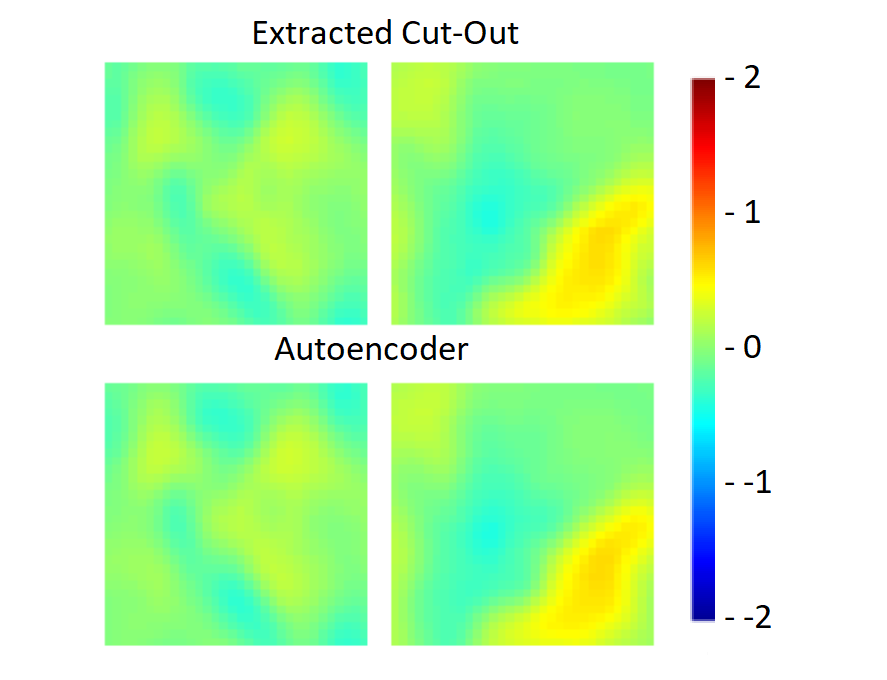}
\caption{Top plots are of random $32^2$ cutouts, and the bottom plots are the same cutouts after the autoencoder encoded and reconstructed them. 
The close similarity between input and decoded 
images indicates excellent performance 
by the autoencoder.
The color bar indicates SSH in units of meters.
} 
\label{fig:6}
\end{figure}

As seen in figure~\ref{fig:6}, the extracted images are very similar to the autoencoder outputs so we conclude that the autoencoder did well in encoding the data
and that a 128-dimension latent vector is sufficient (and possibly too large).
The autoencoder learns how to encode a set of data usually for the purposes of reducing dimensionality by training the ANN to denoise, i.e. ignore insignificant data \citep{baldi2012autoencoders}.

\subsubsection{Normalizing Flow}

\begin{figure}[h!]
\noindent\includegraphics[width=\textwidth]{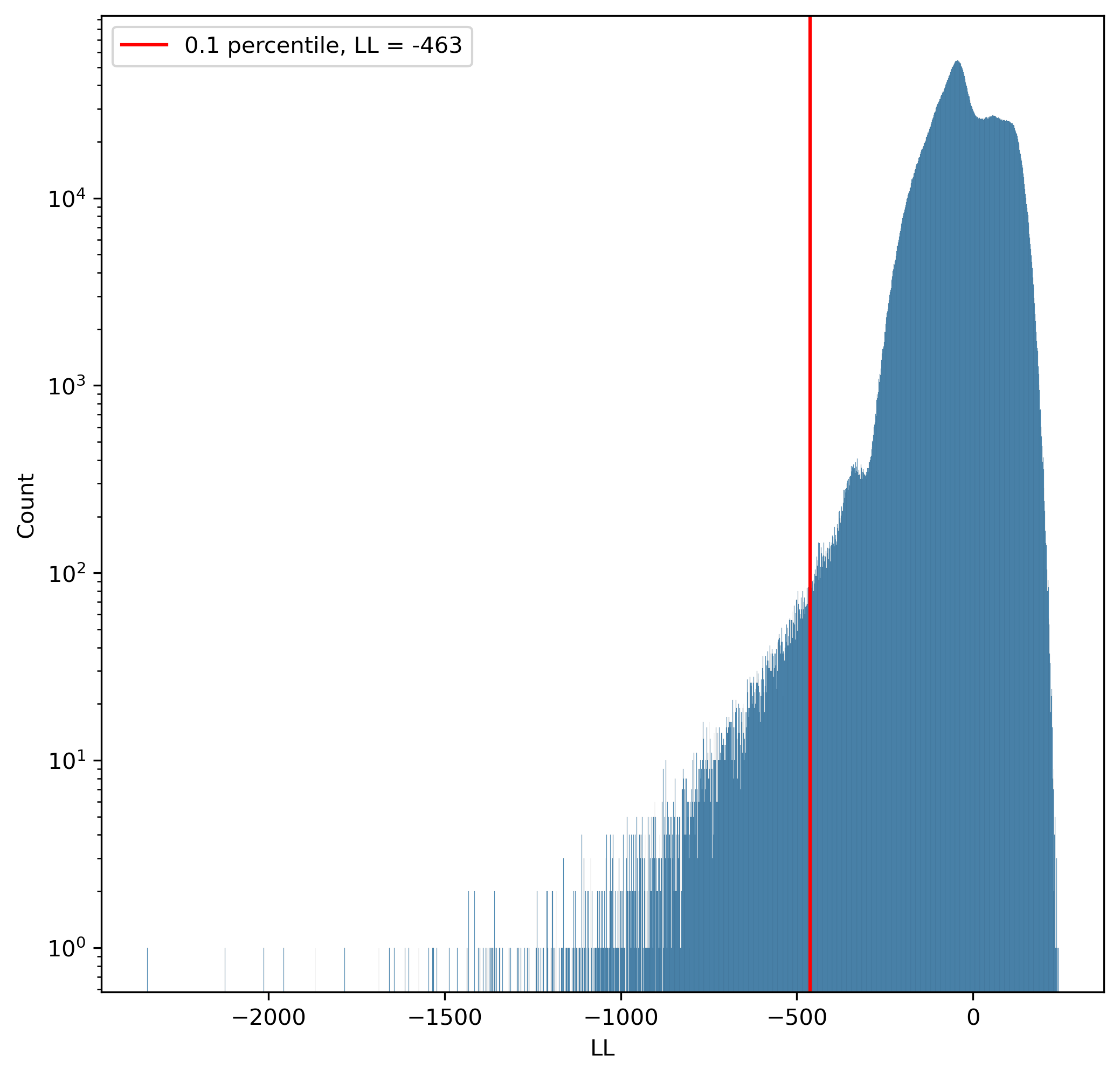}
\caption{Histogram of the Log-Likelihood values for the complete datset.
Lower LL values imply rarer or more unusual cutouts. 
The vertical red line denotes the 0.1 percentile boundary which has an LL value 
of approximately -463. 
Everything to the left of the red line is among the 0.1 percent lowest LL values,
i.e. the anamolies. 
The y-axis has been logged to improve the visibility of the extreme values.} 
\label{fig:7}
\end{figure}

The ``probabilistic'' in probabilistic autoencoder refers to normalizing flow. 
A PAE is based on an autoencoder but interprets probabilistically after training the autoencoder by introducing a normalizing flow \citep{kobyzev2020normalizing}. The normalizing flow takes the 128-dimensional latent distribution of latent vectors and performs transformations on the coordinate mappings to convert the complex latent representation into a one-dimensional 
Gaussian distribution (figure~\ref{fig:7}). 
This takes an uninterpretable representation of that data and makes it more readily understood. Since the data is highly dimensional and complex, performing standard statistical evaluations on the latent vectors would be difficult if not intractable.


\subsubsection{Training and Evaluation}

Training was done on the Nautilus hypercluster using 32Gb of memory and an Nvidia 1080Ti GPU which took approximately four hours for 10 epochs. 
For the hyperparameters, the same architecture as the
PAE model \ulmo of \cite{prochaska2021deep}
was used, with the exception of the latent dimensions
(${\tt n \_ latent}$) which was changed 
from 64 to 128 dimensions.

Training is done in two parts. The first is the development of the autoencoder that maps the extracted cutouts to latent vectors. The second phase takes the latent vectors and transforms them into samples which can be interpreted via Gaussian distribution to estimate the probability of that sample being OOD. Training is done on a random five percent of the extracted images. The remaining images are anlyzed for OOD.




\section{Results}

\subsection{LL-Histogram}

The frequency of occurrence
of the data is quantified by the Log-Likelihood (LL). 
A high LL value means the data is usual or normal where low LL values denote extremes, 
anomalies, and outliers.  For the following we adopt anamolies as those
with the lowest
0.1 percentile of the LL values.

\subsection{Explore Extremes}

\begin{figure}[ht]
\noindent\includegraphics[width=\textwidth]{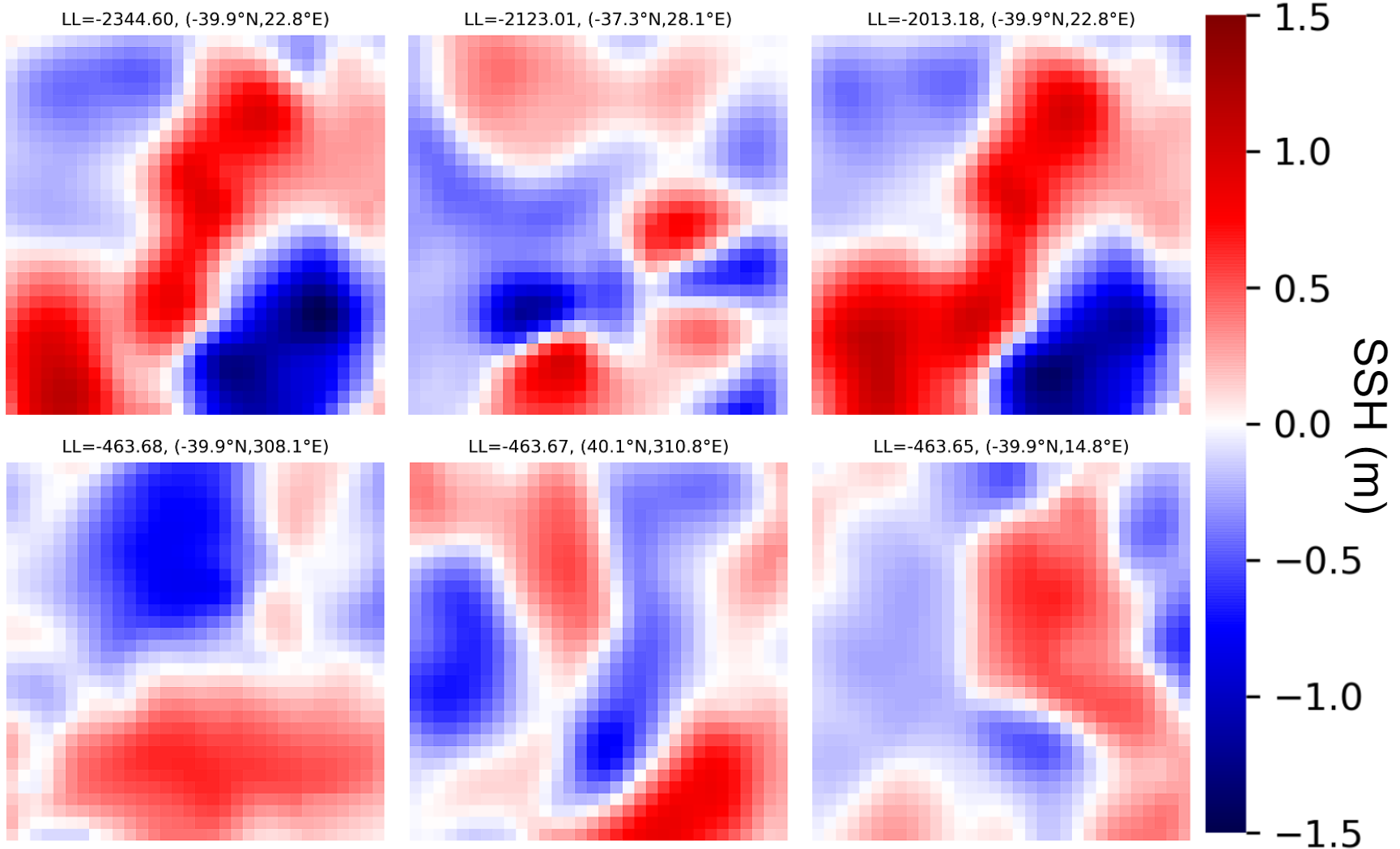}
\caption{Examples of cut-outs that are outliers. The top three images are the most extreme in the dataset going from left to right. The bottom three images are least extreme of the 0.1 percentile, i.e. the least extreme extremes also going from left to right. Coordinates of the cut-outs and LL values are on top of each cut-out.} 
\label{fig:8}
\end{figure}

Figure~\ref{fig:8} depicts examples of images with extreme LL values. 
As stated earlier, SSH values range from approximately -1.5 to 1.5 meters 
above and below the geoid. 
The most extreme outliers have highs and lows that span almost the entire range 
($\Delta \, \rm SSH \approx 3$\,m)
within  one cutout as seen in the top three images 
of figure~\ref{fig:8}. 
The bottom three images of figure~\ref{fig:8} are the least extreme of the
anamolies.
These have $\Delta \, \rm SSH \approx 1.5$\,m
above and below the geoid. 
Images of the most normal values are essentially entirely near zero meters and the plots are basically white squares so they are not shown.

\begin{figure}[ht]
\noindent\includegraphics[width=\textwidth]{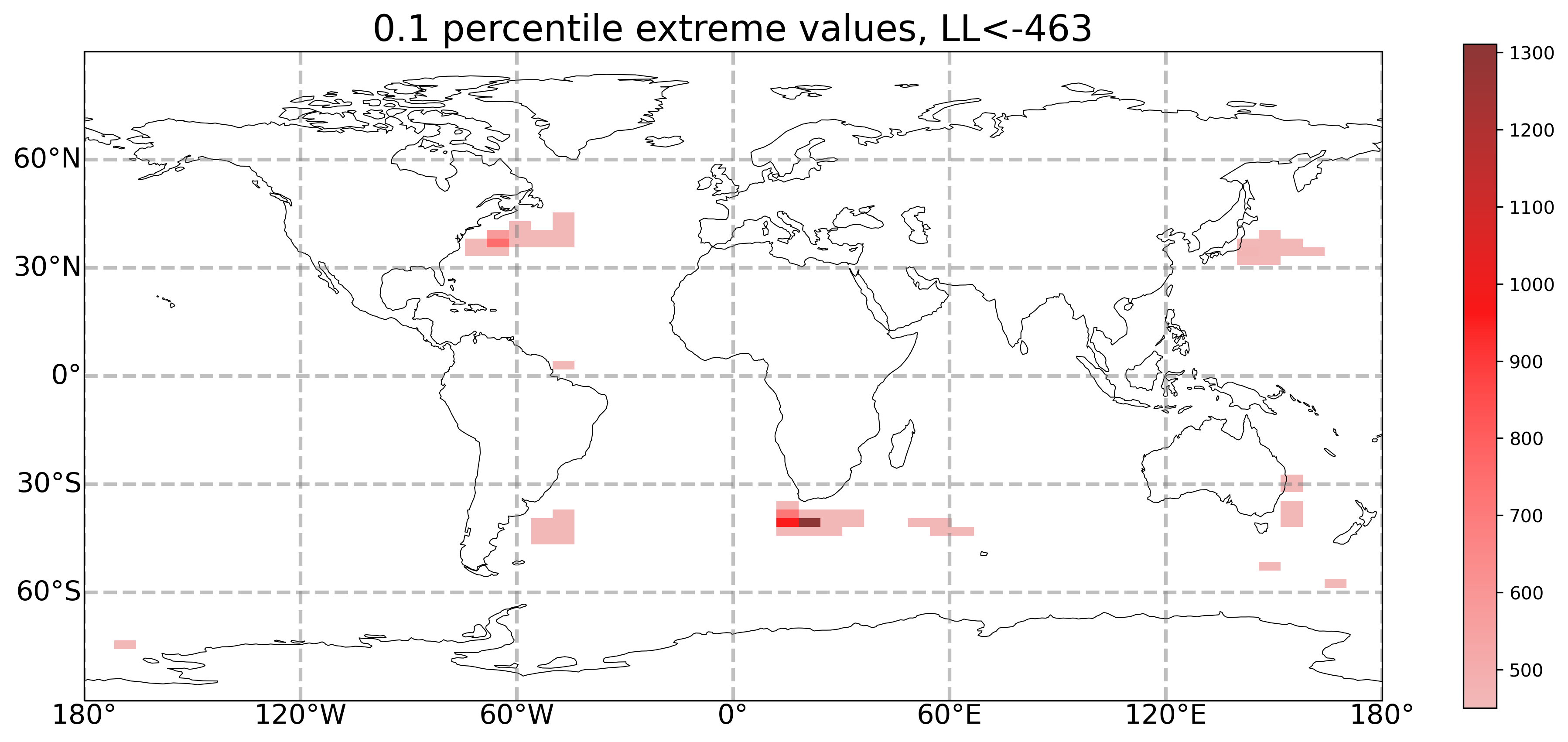}
\caption{
Geographic distribution of the 0.1 percentile anamolies. 
\ulmo\ has picked out all Western-Boundary currents and the Antarctic Circumpolar Current (ACC). These are the Earth's fastest and most dynamic regions. 
The majority of outliers appear in the Gulf Stream and the Agulhas Current.} 
\label{fig:9}
\end{figure}

\begin{figure}[ht]
\noindent\includegraphics[width=\textwidth]{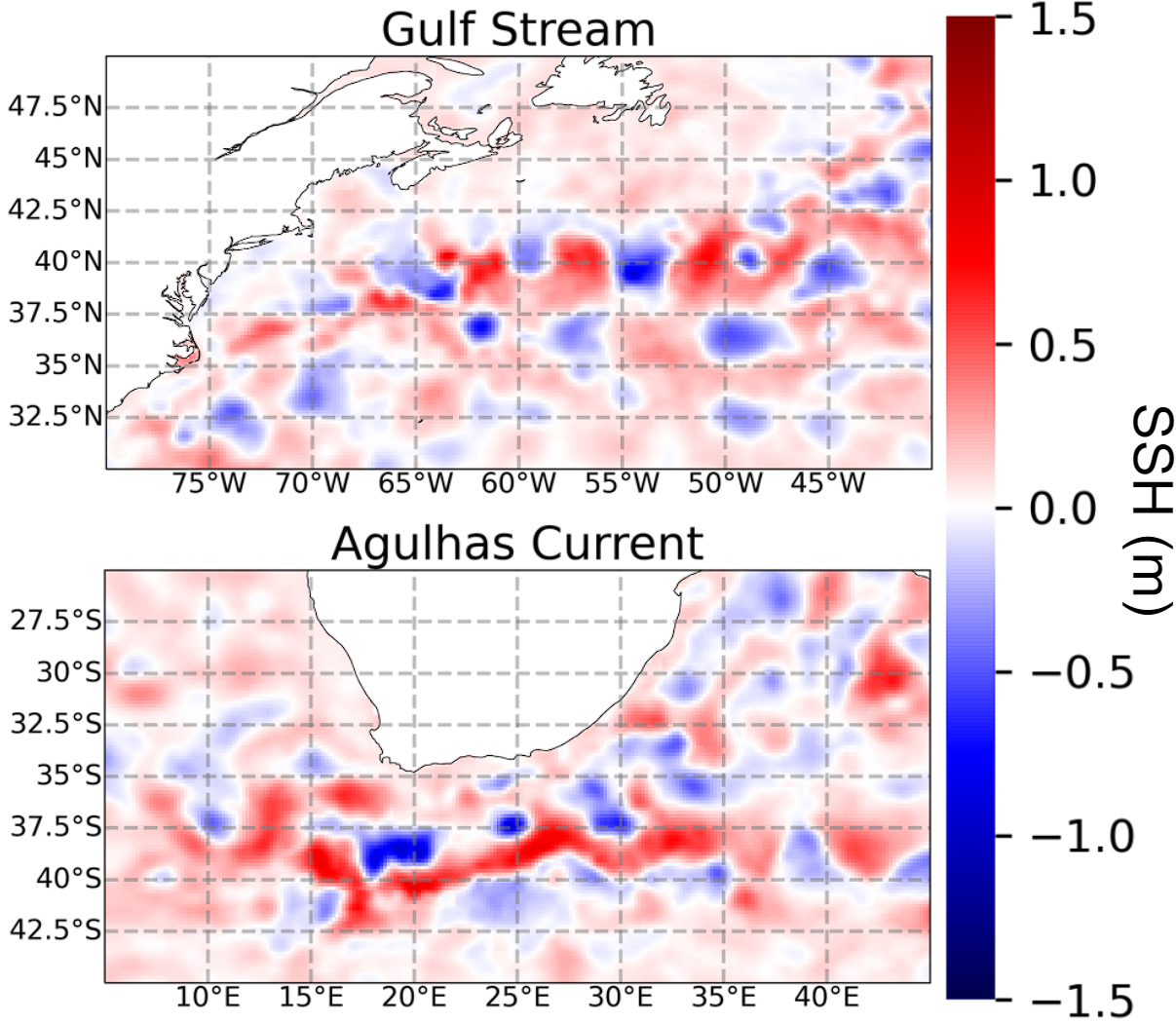}
\caption{Two maps of the raw data where \ulmo\ detected the most anomalies as seen in Figure~\ref{fig:9}. 
These plots are from the first October 2, 1992 time slice. The top plot is the Gulf Stream and the bottom is the Agulhas Current. Between the two, the Agulhas Current has more extremes than the Gulf Stream, and the Agulhas Current extremes are of greater magnitude.} 
\label{fig:10}
\end{figure}

\subsection{Spatial Distribution of Outliers}

Figure~\ref{fig:9} shows the geographic location of the SSH outliers.
There are outliers in all Western-Boundary currents and the Antarctic Circumpolar Current (ACC) particularly near the Ross Sea. The two regions that stand out the most are the Gulf Stream and the Agulhas Current. The top three extremes from figure~\ref{fig:8} all came from the Agulhas Current. Depicted in figure~\ref{fig:10} are two examples,
one each from the Gulf Stream and the Agulhas Current which had the most extreme anomalies.

\section{Discussion and Conclusions}

As can be seen in the examples in figure~\ref{fig:8}, \ulmo\ has detected very strong pressure gradients. Referencing the Geostrophic Balance equations, we can understand that major regions picked out by \ulmo\ are the fastest currents, notably the Gulf Stream and the Agulhas Current. As seen in figure~\ref{fig:8} there are extreme changes in SSH of more than two meters over a distance of about 300 kilometers

  \begin{linenomath*}
  \begin{equation}
   -f \cdot v = -\frac{1}{\rho_\circ} \frac{\partial P}{\partial x} = -g\frac{\partial \eta}{\partial x}
   \end{equation}
   \begin{equation}
    f \cdot u = -\frac{1}{\rho_\circ} \frac{\partial P}{\partial y} = -g\frac{\partial \eta}{\partial y}
  \end{equation}
  \end{linenomath*}
where $f$ is the Coriolis parameter, $\rho_\circ$ is reference density (1027 kg/m$^3$), P is pressure, g is the gravitational constant, and $\eta$ is SSH

The Gulf Stream and the Agulhas Current are both regions of eddy and ring formation which are known to be anomalies in SSH. This is because these regions are where currents collide. The cold Labrador current meets the warm Gulf stream which results in 
an instability of the Gulf Stream resulting in eddies and rings. A similar stronger phenomenon occurs in the ocean by the Cape of Good Hope where the warm Agulhas Current tries to bend around South Africa into the South Atlantic Ocean but is deflected East by the strong cold ACC resulting in Agulhas Current Retroflection. As the Agulhas Current is deflected East it can pinch off and create rings that propagate west.

It is difficult to know whether \ulmo\ has picked out any mesoscale eddies or just regions of strong current-borne turbulent flow. It seems that \ulmo\ has picked out the largest changes in SSH meaning and extreme high next to an extreme low. If they were proper eddies it would be predicted that the features visible in the cutouts would just be lone highs or lows in SSH. It is also possible that \ulmo\ has identified high-low eddy pairs akin to a binary star system in space. Another possibility is that \ulmo\ did not pick-out eddies because they range vertically on the order of tens of centimeters above or below the geoid.

As far as future work is concerned, It may be worthwhile to see what \ulmo\ identifies with 64$^2$ pixel grid extractions. We also 
encourage application of \ulmo\ to future SSH datasets like those produced by the upcoming Surface Water and Ocean Topography (SWOT) mission developed by NASA, Centre National D'Etudes Spatiales (CNES) and the Canadian Space Agency (CSA) and United Kingdom Space Agency. The SWOT mission will record SSH measurements at an unprecedented 0.05 degree scale which will be the first SSH dataset in which submesoscale features can be resolved.

\bibliographystyle{copernicus}
\bibliography{agusample.bib}

\end{document}